\documentclass[11pt,twoside]{article}
\usepackage{./asp2014}

\aspSuppressVolSlug
\resetcounters

\begin{document}

\title{Binaries with a $\delta$~Scuti component: Results from a long term observational survey, updated catalogue and future prospects}
\author{Alexios Liakos$^1$ and Panagiotis Niarchos$^2$
\affil{$^1$National Observatory of Athens, Institute for Astronomy, Astrophysics, Space Applications and Remote Sensing, Penteli, Athens, Hellas; \email{alliakos@noa.gr}}
\affil{$^2$National \& Kapodistrian University of Athens, Department of Astrophysics, Astronomy and Mechanics, Zografos, Athens, Hellas; \email{pniarcho@phys.uoa.gr}}}

\paperauthor{Alexios Liakos}{alliakos@noa.gr}{}{National Observatory of Athens}{Institute for Astronomy, Astrophysics, Space Applications and Remote Sensing}{Athens}{}{GR15236}{Hellas}
\paperauthor{Panagiotis Niarchos}{pniarcho@phys.uoa.gr}{}{National \& Kapodistrian University of Athens}{Department of Astrophysics, Astronomy and Mechanics}{Athens}{}{GR15784}{Hellas}

\begin{abstract}
Results from a six year systematic observational survey on candidate eclipsing binaries with a $\delta$~Sct component are presented. More than a hundred systems with component(s) of A-F spectral types were observed in the frame of this survey in order to be checked for possible pulsational behaviour. The $\sim14\%$ (13 cases) of the currently known such systems were discovered during this survey. Using all the available information from the literature, an updated list with all the currently known systems of this type is presented, while possible correlations between their pulsational and binarity properties are discussed.

\end{abstract}

\section{Motivation for studying and history of these systems}

On one hand, binary systems and especially the eclipsing ones (hereafter EBs) can be used as essential tools for the calculation of stellar absolute parameters. On the other hand, asteroseismology provides the means for understanding the stellar interior and evolution using the pulsations as signals. Thus, the combination of all the information coming from binaries with pulsating components is very useful for the complete study of evolution of interacting stars.

\citet{MK02,MK04} introduced the oEA (oscillating EA) stars as the (B)A-F spectral type mass-accreting MS pulsators in semi-detached Algol-type eclipsing binaries. \citet{SO06a} noticed for first time the connection between pulsation and orbital periods for EBs with a $\delta$~Sct member based on a sample of 20 systems. \citet{SO06b} published a catalogue with confirmed and candidate such systems. \citet{LI12} published an updated catalogue with 75 confirmed systems with a $\delta$~Sct companion and derived new correlations between their fundamental stellar characteristics. \citet{ZH13} presented the first theoretical approach for the connection between orbital and dominant pulsation periods.

\section{The long term survey and data analysis methods}

The candidate systems were selected from the catalogue of \citet{SO06b}. The survey started on 2006 and ended on 2012, while $\sim$550 nights ($\sim$2200 hrs) were spent for observations. Photometric observations were made using the telescopes (40~cm, 25~cm, 20~cm) of the Gerostathopoulion Observatory of the University of Athens and the 1.2~m telescope of the Kryonerion Astronomical Station of the National Observatory of Athens. Few spectroscopic observations were made at Skinakas Observatory with the 1.3~m telescope. 108 candidate systems were observed in total. 13 of them were found to be new cases of EBs with $\delta$~Sct component, while 8 of them are still ambiguous (i.e. the possible pulsations were not detected due to instrumentation limits). In addition, we observed systematically 9 known oEA systems for better estimation of their pulsation properties. Summarizing the above, complete multi-colour light curves (hereafter LCs) and systematic observations (i.e. in a time span of weeks/months) were obtained for 21 systems with $\delta$~Sct component.

The differential aperture photometry method was applied in the photometric data. For the cases for which spectroscopic observations had been made, the \textit{Broadening Functions} method \citep{RU02} was applied for extracting the radial velocities (hereafter RVs) of the components. Spectra taken inside the eclipses were compared with other of standard stars and were used to find the spectral type of one of the components (typically that of the primary). The binary model based on the LCs and, in a few cases, also on the RVs was made using the \textsc{PHOEBE} 0.29d software \citep{PZ05}. Using this model, the absolute parameters of the system were calculated and the evolutionary status of its components was estimated. For systems that show orbital period modulations, $O-C$ diagram (Eclipse Time Variation--ETV) analysis was also performed using a MatLab code \citep{ZA09}. Then, the binary model was subtracted from the observed points and the LC residuals, except for those inside the eclipses, were further analysed for pulsation presence using the \textsc{PERIOD04} v.1.2 software. Once pulsation properties were found (i.e. frequencies, amplitudes, phases), they were put as inputs in the \textsc{FAMIAS} software \citep{ZI08} in order to identify the pulsation modes.

Finally, detailed description for this survey and its results can be found in the following papers: \citet{LN09, LN12, LN13, LI12}, while the results are also available online\footnote{\url{http://alexiosliakos.weebly.com/binaries-with-a-delta-sct-member.html}}.

\section{Updated catalogue}

Using the available information from literature, all the currently known systems with a $\delta$~Sct component were collected and they are presented in Table 1. The catalogue includes the name of the system, its geometrical configuration, its orbital period ($P_{\rm orb}$), the dominant pulsation frequency $f$ and the absolute parameters (mass, radius) of the $\delta$~Sct star. In fig.~\ref{stat-pp} the statistics for these systems is also presented.

\begin{table}
\caption{The new catalogue of binaries with a $\delta$~Sct component.}
\smallskip
\begin{center}
{\small
\scalebox{0.75}{
\begin{tabular}{lcc ccc lcc ccc}  
\tableline
\noalign{\smallskip}
Name             & Type &$P_{\rm orb}$& $f$ &  $M$  & $R$   &        Name        &  Type &$P_{\rm orb}$&  $f$   &  $M$  &   $R$ \\
                 &      &    (days)   &(c/d)&(M$_{\odot})$ &(R$_{\odot}$)&       &       &    (days)   & (c/d)  &(M$_{\odot})$ &(R$_{\odot}$)\\
\tableline
Aqr CZ	        &	SD	&	0.86275	&35.508	&	2.0	&	1.9	&	Her V0644	     &U$^{1}$ &	11.85859&	8.688	&	1.5	&		\\
Aqr DY	        &	SD	&	2.15970	&23.370	&	1.8	&	2.1	&	Her V0994	     &	U	  &	2.08309	&	10.563	&		&		\\
Aql QY	        &	SD	&	7.22954	&10.656	&	1.6	&	4.1	&	HIP 7666	     &	D	  &	2.37232	&	24.450	&		&		\\
Aql V1464	    &	SD	&	0.69777	&24.621	&	2.2	&	2.5	&	Hor TT	         &	U	  &	2.60820	&	38.700	&		&		\\
Aql V729	    &	SD	&	1.28191	&28.034	&	1.5	&	2.0	&	Hya AI	         &	D	  &	8.28970	&	7.246	&	2.0	&	1.8	\\
Aur KW (14)	    &U$^{1}$&   3.78900 &11.429	&	2.3	&	4.0	&	Hya KZ	         &U$^{3}$ &	9782.00 &	16.807	&		&		\\
Aur V551	    &	D	&	1.17320	&7.727	&		&		&	Hya RX	         &	SD	  &	2.28170	&	19.380	&	1.7	&	1.7	\\
Boo EW	        &	SD	&	0.90630	&48.008	&	1.4	&	1.7	&11754974$^{c}$	     &U$^{1}$ &	343.000	&	16.342	&	1.5	&		\\
Boo YY	        &	SD	&	3.93307	&16.318	&	2.0	&	1.9	&10661783$^{c}$	     &	SD	  &	1.23136	&	28.135	&		&		\\
Cam Y	        &	SD	&	3.30570	&17.065	&	1.7	&	2.9	&3858884$^{c}$	     &	D	  &	10.0486	&	7.231	&	1.9	&	3.1	\\
Cap TY	        &	SD	&	1.42346	&24.222	&	2.0	&	2.5	&4544587$^{c}$	     &	D	  &	2.18909	&	48.022	&	2.0	&	1.8	\\
Cas AB	        &	SD	&	1.36690	&17.153	&	2.3	&	2.0	&	Lac AU	         &	SD	  &	1.39259	&	58.217	&	2.0	&	1.8	\\
Cas IV	        &	SD	&	0.99852	&32.692	&	2.0	&	2.1	&	Leo DG	         &	U	  &	4.14675	&	11.994	&	2.0	&	3.0	\\
Cas RZ	        &	SD	&	1.19530	&64.197	&	2.0	&	1.6	&	Leo WY	         &	SD	  &	4.98578	&	15.267	&	2.3	&	3.3	\\
Cas $\beta$	    &U$^{1}$&	27.0000	&9.911	&		&		&	Leo Y	         &	SD	  &	1.68610	&	34.484	&	2.3	&	1.9	\\
Cep XX	        &	SD	&	2.33732	&32.258	&	1.9	&	2.1	&	Lep RR	         &	SD	  &	0.91543	&	33.280	&	1.8	&	2.2	\\
Cet WY	        &	SD	&	1.93969	&13.211	&	1.7	&	2.2	&	Lyn CL	         &	SD	  &	1.58606	&	23.051	&	1.8	&	2.4	\\
Cha RS	        &	D	&	1.66987	&11.628	&	1.9	&	2.2	&	Lyn CQ	         &	U	  &	12.5074	&	8.868	&		&		\\
CMa R	        &	SD	&	1.13590	&21.231	&	1.7	&	1.8	&	Lyn SZ	         &U$^{1}$ &	1181.10	&	8.299	&		&		\\
105906206$^{a}$	&	D	&	3.69457	&9.417	&	2.3	&	4.2	&	Mic VY	         &	SD	  &	4.43637	&	12.234	&	2.4	&	2.2	\\
CVn 4	        &U$^{1}$&	124.440	&8.595	&		&		&	Oph V0577	     &	D	  &	6.07910	&	14.388	&	1.7	&	1.8	\\
Cyg UW	        &	SD	&	3.45078	&27.841	&	1.9	&	2.2	&	Oph V2365 	     &	SD	  &	4.86560	&	14.286	&	2.0	&	2.2	\\
Cyg V0346	    &	SD	&	2.74330	&19.920	&	2.3	&	3.8	&	Ori EY	         &	U	  &	16.7878	&	9.709	&		&		\\
Cyg V0469	    &	SD	&	1.31250	&35.971	&	3.3	&	2.7	&	Ori FL	         &	SD	  &	1.55098	&	18.178	&	2.9	&	2.1	\\
Del BW	        &	SD	&	2.42319	&25.100	&	1.5	&	2.2	&	Ori FO	         &	U	  &	18.8006	&	34.247	&		&		\\
Del $\delta$	&U$^{2}$&	40.580	&6.378	&		&		&	Ori FR	         &	SD	  &	0.88316	&	38.600	&		&		\\
Dra GK	        &	D	&	16.960	&8.790	&	1.8	&	2.8	&	Ori V1004	     &U$^{1}$ &	2.74050	&	15.365	&		&		\\
Dra HL	        &	SD	&	0.94428	&26.914	&	2.5	&	2.5	&	Pav MX	         &	SD	  &	5.73084	&	13.227	&		&		\\
Dra HN	        &	D	&	1.80075	&8.558	&		&		&	Peg BG	         &	SD	  &	1.95267	&	25.544	&	2.5	&	3.0	\\
Dra HZ	        &	D	&	0.77294	&51.068	&	3.0	&	2.3	&	Peg GX	         &U$^{1}$ &	2.34100	&	17.857	&		&		\\
Dra OO 	        &	D	&	1.23837	&41.867	&	2.0	&	2.0	&	Peg IK	         &U$^{1}$ &	21.7240	&	22.727	&		&		\\
Dra SX	        &	SD	&	5.16957	&22.742	&	1.8	&	2.3	&	Per AB	         &	SD	  &	7.16030	&	5.107	&	1.9	&	2.0	\\
Dra TW	        &	SD	&	2.80690	&17.986	&	2.2	&	2.6	&	Per IU	         &	SD	  &	0.85700	&	43.131	&	2.4	&	1.9	\\
Dra TZ	        &	SD	&	0.86603	&50.993	&	1.8	&	1.7	&	Pup HM	         &	U	  &	2.58972	&	31.900	&		&		\\
Dra WX	        &	U	&	1.80186	&35.468	&		&		&	Pyx XX	         &	D	  &	1.15000	&	38.110	&		&		\\
Eri AS	        &	SD	&	2.66410	&59.172	&	1.9	&	1.6	&	Ser AO	         &	SD	  &	0.87930	&	21.505	&	2.5	&	1.7	\\
Eri TZ	        &	SD	&	2.60610	&18.718	&	2.0	&	1.8	&	Sge UZ	         &	SD	  &	2.21574	&	46.652	&	2.1	&	1.9	\\
Gru RS	        &	U	&	11.500	&6.803	&		&		&	Tau AC	         &	SD	  &	2.04340	&	17.535	&	1.5	&	2.3	\\
3889-0202$^{b}$	&	SD	&	2.71066	&22.676	&		&		&	Tau $\rho$	     &U$^{1}$ &	460.700	&	14.925	&	1.2	&		\\
4293-0432$^{b}$	&	SD	&	4.38440	&8.000	&		&		&	Tau $\theta^{2}$ &	D	  &	140.728	&	13.228	&	2.9	&		\\
4588-0883$^{b}$	&	SD	&	3.25855	&20.284	&		&		&	Tau V0777 	     &U$^{1}$ &	5200.00	&	5.486	&		&		\\
HD 061199	    &	U	&	3.57436	&25.257	&		&		&	Tel IZ	         &	SD	  &	4.88022	&	13.558	&		&		\\
HD 062571	    &	SD	&	3.20865	&9.051	&		&		&	Tri X	         &	SD	  &	0.97151	&	45.455	&		&		\\
HD 099612	    &	D	&	2.77876	&14.714	&		&		&	Tuc $\theta$	 &U$^{2}$ &	7.10360	&	15.946	&	2.0	&		\\
HD 172189	    &	D	&	5.70165	&19.608	&	2.1	&	4.0	&	UMa IO	         &	SD	  &	5.52017	&	22.015	&	2.1	&	3.0	\\
HD 207651	    &	U	&	1.47080	&15.434	&		&		&	UMa VV	         &	SD	  &	0.68738	&	51.299	&	2.5	&	1.8	\\
HD 220687	    &	D	&	1.59425	&26.169	&		&		&	UNSW-V-500	     &	SD	  &	5.35048	&	13.624	&	1.5	&	2.4	\\
HD 50870	    &U$^{1}$&	     	&17.162	&		&		&0975-17281677$^{d}$ &	U	  &	3.01550	&	18.702	&		&		\\
HD 51844	    &U$^{2}$&	33.4983	&12.213	&	2.0	&	3.5	&1200-03937339$^{d}$ &	SD	  &	1.17962	&	30.668	&	1.5	&	2.2	\\
Her BO	        &	SD	&	4.27281	&13.430	&	1.8	&	2.5	&	Vel AW	         &	U	  &	1.99245	&	15.200	&		&		\\
Her CT	        &	SD	&	1.78640	&52.937	&	2.3	&	2.1	&	Vel BF	         &	SD	  &	0.70400	&	44.940	&	2.0	&	1.8	\\
Her EF	        &	SD	&	4.72920	&10.070	&	1.8	&	2.8	&	Vir FM	         &U$^{2}$ &	38.3240	&	13.908	&		&		\\
Her LT	        &	SD	&	1.08404	&30.800	&		&		&	Vul 18	         &U$^{2}$ &	9.31000	&	8.230	&		&		\\
Her TU	        &	SD	&	2.26690	&17.986	&		&		&                    &        &         &           &       &       \\
\tableline
\noalign{\smallskip}
\multicolumn{12}{l}{Catalogues: (a) CoRoT, (b) Guide Star Catalogue (GSC), (c) Kepler, (d) USNO-A2.0, D=Detached, SD=Semi-detached,}    \\
\multicolumn{12}{l}{U=Unknown, $^{1}$Single-line; $^{2}$Double-line Spectroscopic Binary, $^{3}$Variation detected through O-C analysis}\\
\end{tabular}}}
\end{center}
\end{table}

\section{Correlations}

As discussed in many papers \citep{SO06a, LI12, ZH13} there is a correlation between $P_{\rm orb}$-$P_{\rm puls}$. The common thing in these papers is that their samples include cases with relatively small orbital period values. However, many systems with much larger periods have been discovered. In Fig.~\ref{stat-pp} we have plotted all the currently known systems. It seems that the linear correlation between $P_{\rm orb}$-$P_{\rm puls}$ stops near the value of $P_{\rm orb}\sim13$~days, and after that these quantities are uncorrelated. Therefore, the following correlations concern only the systems with $P_{\rm orb}<13$~days.

\articlefiguretwo{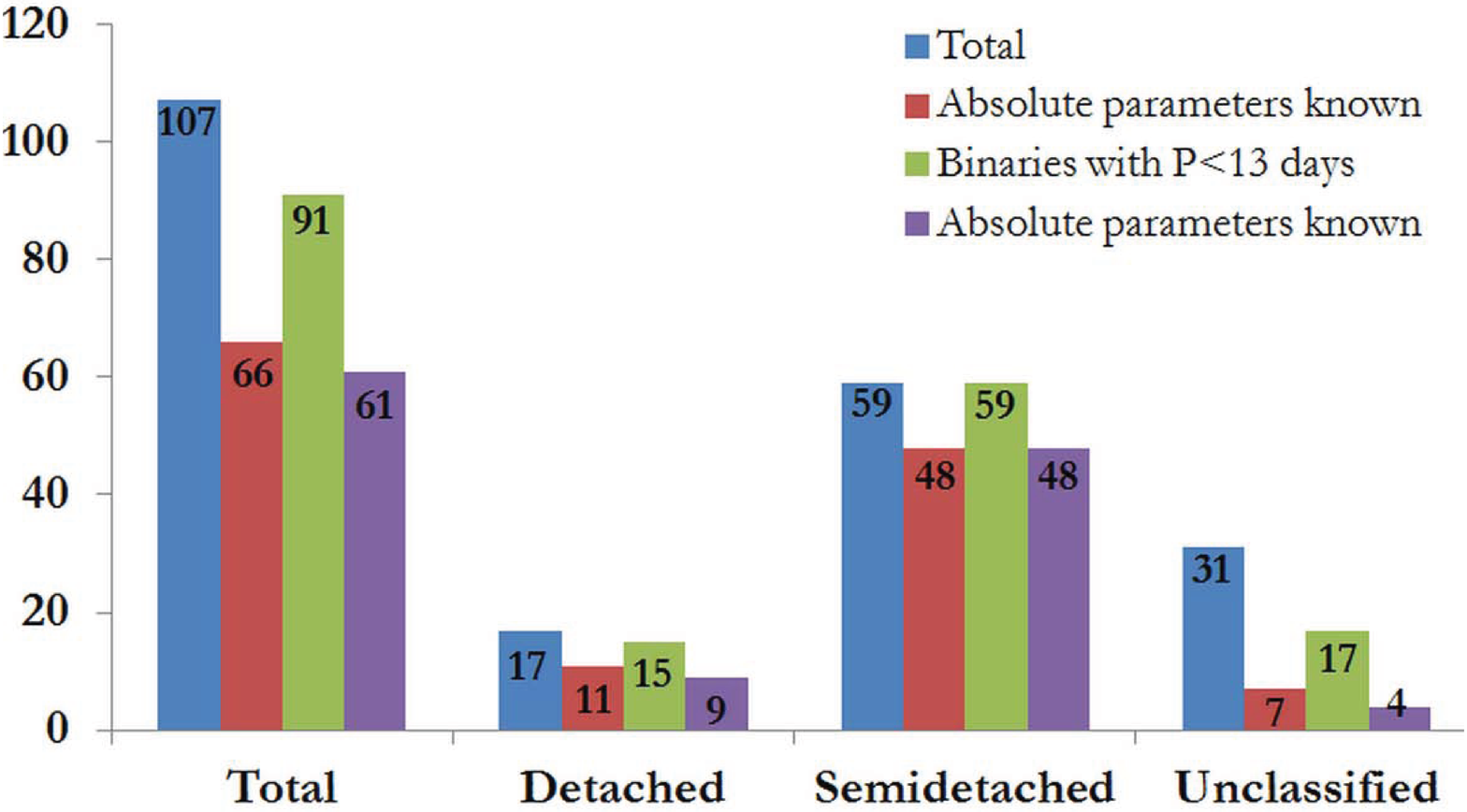}{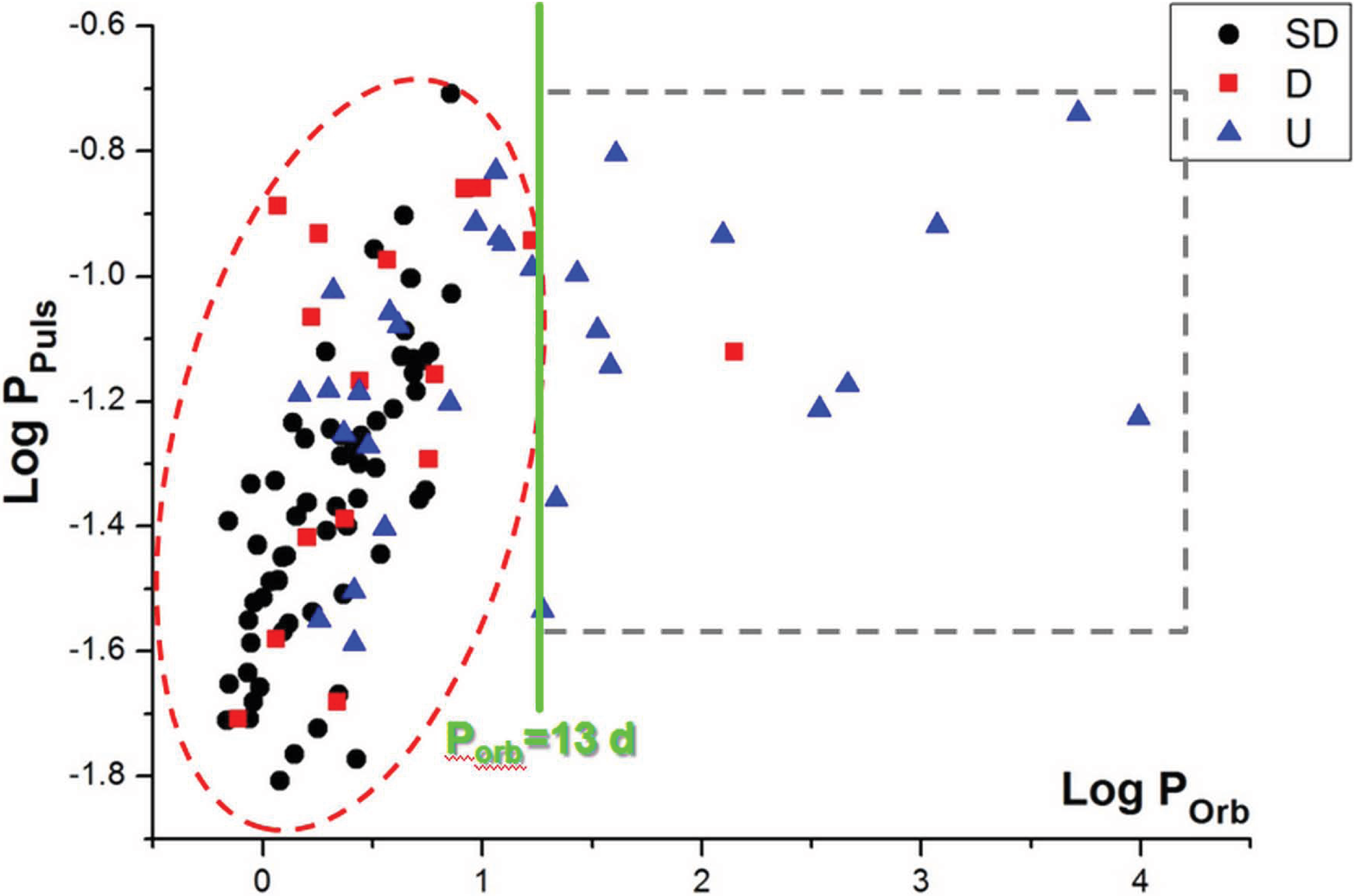}{stat-pp}{\emph{Left:} Statistics of binaries with a $\delta$~Sct component. \emph{Right:} $P_{\rm orb}$-$P_{\rm puls}$ plot for binaries with a $\delta$~Sct component. The subsets have been distinguished before and after the critical orbital period value of 13 days. For the systems included in the elliptical region ($P_{\rm orb}<13$~days) there is clear correlation between $P_{\rm orb}$-$P_{\rm puls}$, while for those included in the rectangular region there is no such correlation.}

In Fig.~\ref{ppfit_all} all the available data points (i.e. $P_{\rm orb}$-$P_{\rm puls}$ values for the 91 so far known systems with $P_{\rm orb}<13$~days) are presented along with the linear fits as given by various researchers. In particular, the (empirical) fit of \citet{SO06a} is based on 20 systems, the (empirical) fit of \citet{LI12} on 70, the (theoretical) fit of \citet{ZH13} on 69, while the one of the present work on 91. The new correlation for these quantities based on the current sample is the following: $\log P_{\rm puls}=0.56 (1) \log P_{\rm orb}-1.52(2)$


\articlefigure[width=.5\textwidth]{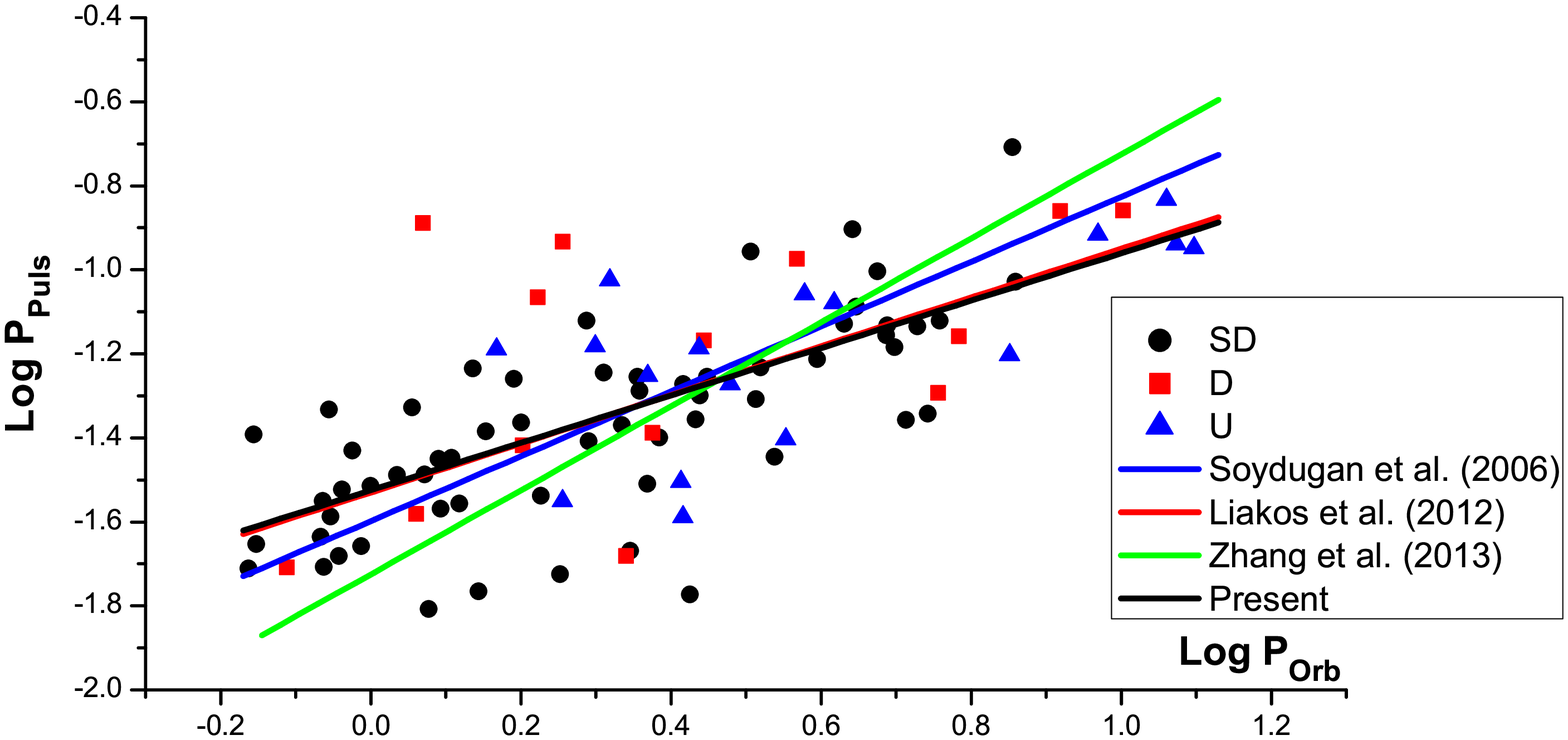}{ppfit_all}{Various linear fits on the $P_{\rm orb}$-$P_{\rm puls}$ values of all kind of systems based on different samples (see text for details).}

In Fig.~\ref{gmr} (right part) the pulsating components of these systems are shown within the $M-R$ diagram showing that in most of them the pulsations begin when the star is still inside the Main Sequence band, in contrast with the simple $\delta$~Sct stars. In the left part of the same figure, the $\delta$~Sct stars of these systems, whose their absolute parameters have been calculated, are presented within the $\log g-P_{\rm puls}$ diagram. Along with them and for comparison reasons, the linear fit of \citet{LI12} based on 46 systems, the present one based on 59 systems and the one for the single $\delta$~Sct stars as given by \citet{CL90} are also presented. Finally, we found that the $P_{\rm puls}$ of the $\delta$~Sct star in a binary system is correlated with its evolutionary stage according to the following (empirical) relation: $\log g=-0.5(1) \log P_{\rm puls} + 3.4(2)$


\articlefiguretwo{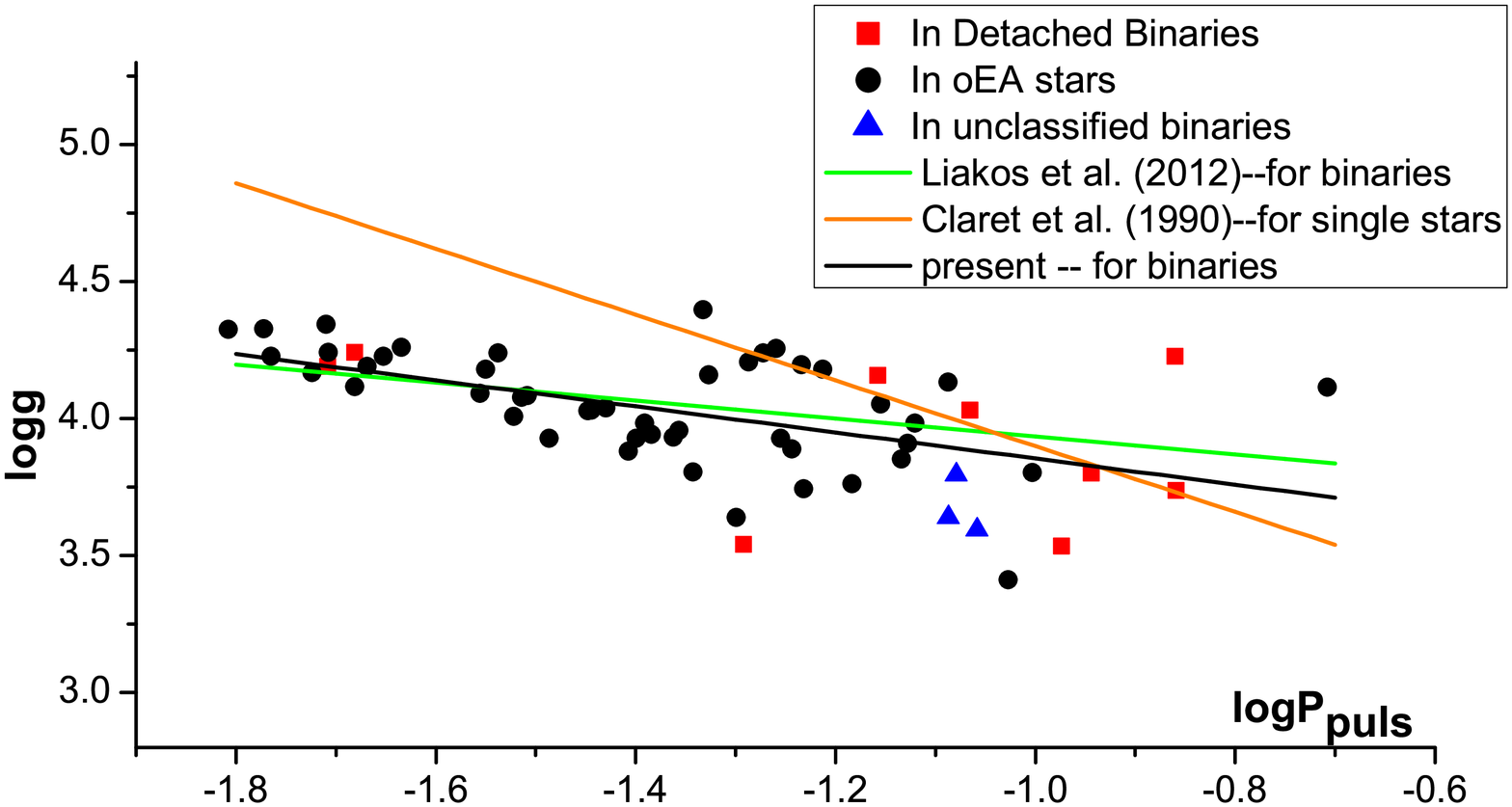}{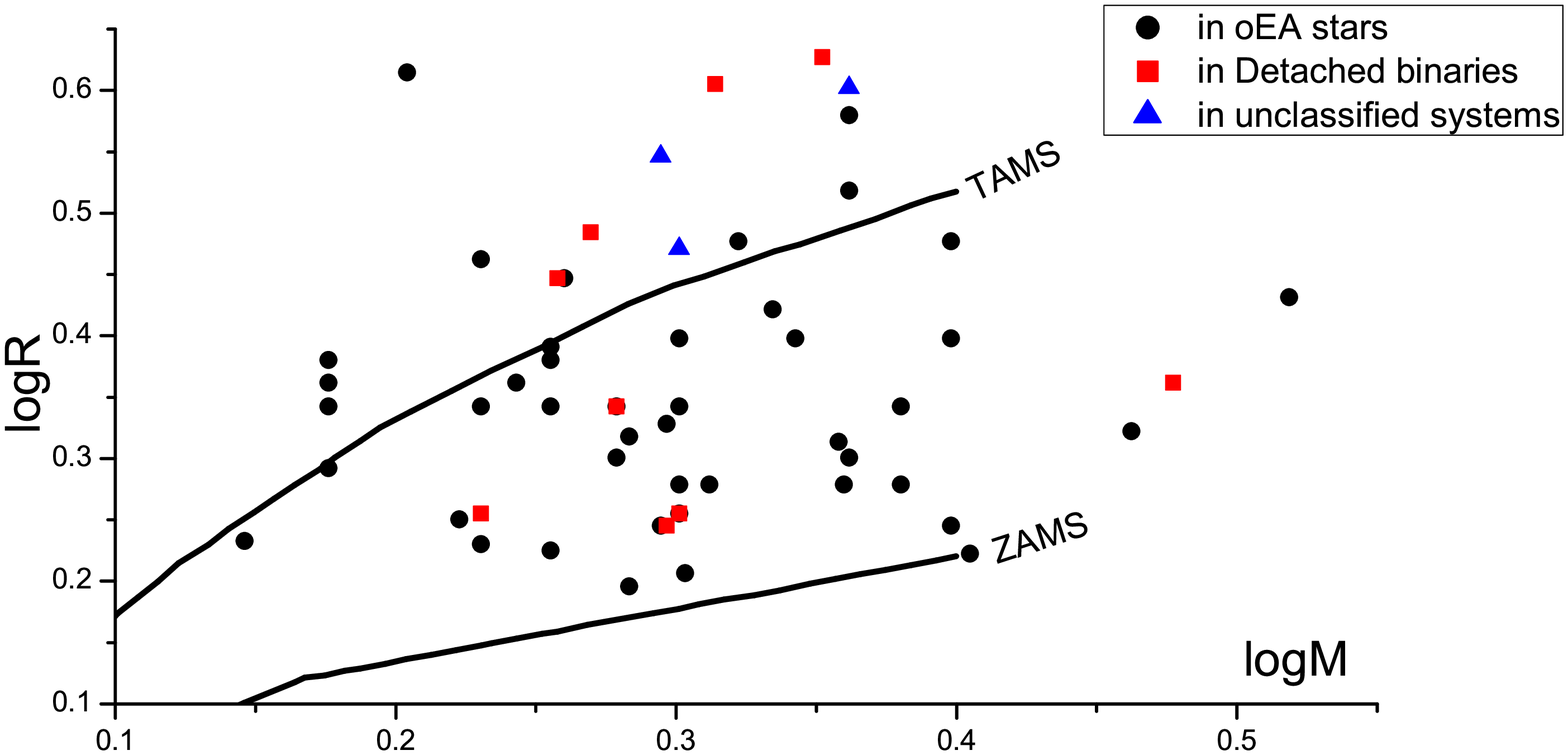}{gmr}{The $\log g-P_{\rm puls}$ various linear fits (left) for the $\delta$~Sct stars in binaries (see text for details) and their location within the $M-R$ diagram (right).}

\section{Discussion and future prospects}

We can plausibly conclude that there is certain correlation between $P_{\rm orb}$-$P_{\rm puls}$, however, it seems that binarity does not affect the pulsation properties in systems with $P_{\rm orb}>$13 days. $\delta$~Sct stars in binaries are mostly Main Sequence stars and evolve differently than the single ones.

Theoretical establishment for all correlations is needed, while surveys for discovering new systems of this kind are highly recommended to enrich the current sample. Given that mass transfer procedures during the binaries' life affect their evolution, continuous monitoring for several decades for cases with rapid mass transfer is proposed in order to obtain conclusions about the mass transfer influence on pulsations.

The 2.3~m Aristarchos telescope (Helmos Observatory) of the National Observatory of Athens has recently joined the effort for observations of these systems, while a new instrument, namely \textit{Aristarchos Wide Field Camera} (AWFC), will produce a field of view of $\sim30'\times30'$ in a CCD size of 4K$\times$4K (end of 2015), and it is expected that it will help a lot the observations, especially in star-poor fields.

\acknowledgements This work was performed in the framework of PROTEAS project within GSRT's KRIPIS action, funded by Hellas and the European Regional Development Fund of the European Union under the O.P. Competitiveness and Entrepreneurship, NSRF 2007-2013.




\end{document}